# Giant chiroptical effect caused by the electric quadrupole

Tong Wu, Weixuan Zhang, Rongyao Wang, and Xiangdong Zhang[*]

Beijing Key Laboratory of Nanophotonics & Ultrafine Optoelectronic Systems, School of Physics, Beijing Institute of Technology, 100081, Beijing, China.

*Correspondence to zhangxd@bit.edu.cn

ABSTRACT Recently, there is a great interest in studying ultrasensitive detection and characterization of biomolecules using plasmonic particles, because they are of considerable importance in biomedical science and pharmaceutics. So far, all the theories on plasmon-induced circular dichroism (CD) are based on the dipole approximation, the electric quadrupolar contribution is generally considered to be relatively small and neglected. Here we demonstrate that the electric quadrupolar contribution not only cannot be ignored, but also plays a key role in many cases. Particularly, for the chiral medium that possesses preferential molecular orientations and is located at the hotspot of plasmonic nanostructures, the plasmonic CD strength contributed by molecular electric quadrupoles (EQ) can be two orders of magnitude higher than that contributed by molecular electric/magnetic dipoles. Unlike the case of dipole approximation, molecular EQ associated plasmonic CD activity appears mainly at the plasmonic resonance absorptions that facilitate the optically enhanced near-field with steep electric filed gradients, and is correlated with the boosted emission rate of a molecular EQ. Based on such physical understandings, we can design nanostructures to realize giant chiroptical effect using the EQ contribution according to the requirements, which provide a new strategy for ultrasensitive detection and quantification of molecular chirality.

Detection and quantification of chiral enantiomers of biomolecules are of considerable importance for biomedical diagnostics and pathogen analyses[1-2]. Circular dichroism (CD) spectroscopy is generally used for analyzing molecular chirality, which typically measures the small differences in the interaction of left- and right-circularly polarized light with a chiral substance[1-3]. However, due to inherent weak light–molecule interactions involved in the chiroptical phenomena, chiral analyses by those conventional CD spectroscopic methods are restricted to a relatively high concentration of analytes. This is frequently an impediment for a practical use because an ultrasensitive probe of tiny amounts of chiral substance is highly demanded for practical biosensing applications in biomedical and pharmaceutical fields.

Recently, intense theoretical and experimental studies have been devoted to plasmon enhanced CD activity, because a weak molecular CD signal in the UV spectral region can be both enhanced and transferred to the visible/near-infrared region when chiral molecules are adsorbed at the surfaces of metallic nanoparticles (NPs) or in the nanogaps (i.e. hot spots) of particles' clusters[4-16]. In the previous theoretical works [8,9,11,16], chiral molecules are usually treated as a combination of electric dipole (ED) and magnetic dipole (MD). The electric quadrupolar contribution is always ignored, because it is generally believed that the role of the electric quadrupole (EQ) is much smaller than that of the dipole[3].

However, recent investigations have shown that the EQ plays an important role in some other optical activity phenomena[17-19]. For example, the interaction between molecular electric dipole and electric quadrupole gives a large contribution to the Raman optical signal of chiral molecules[17-18]. Recent experiments have also shown that an unexpectedly large CD response, far beyond the theoretical prediction, can be observed, when the colloidal silver nanocubes (NC) are capped by chiral molecules[4,12]. This strong CD signal is also sensitive to the direction of the chiral molecule relative to a NC surface. A latest work has found that inversion of the chiral molecule's orientation with respect to the surface of NC led to inversion of the plasmonic CD signal[13], which cannot be understood by the previous theories based on the dipole approximation.

Motivated by above investigations, in this work we present an exact T-matrix method to calculate the CD of molecule−NP nanocomposites with the EQ contribution. On the basis of such a method, we study the EQ contribution to the plasmon-induced CD signals for the single nano particle, dimer and trimer. The above unsolved experimental phenomena can be understood very well. We find when the chiral molecule is inserted into the hotspot of a nano dimer (trimer), extremely strong CD signals caused by the molecular electric quadrupole can be observed. These strong CD signals only occur at the plasmonic resonance frequency where steep electric filed gradients are induced. Furthermore, we also find the unexpected large CD signals are also related with the boosted emission rate of an electric quadrupole in the hotspot[20-21] and the symmetry of the nanostructures [22-24]. We can design nanostructures to obtain strong CD signals by using the EQ contribution.

**Results**

**Enhanced chiroptical effect by a nanoparticle.** We consider a hybrid system consisting of a chiral molecule and a silver NP as shown in the inset of Fig. 1a, which is excited by left- or right- handed circularly polarized lights with the angular frequency $\omega$. The chiral molecule is located at the origin of the coordinate, and the NP with radius 10 nm is put at $R_0 = (0,\ 0,\ -11\,\text{nm})$. The chiral molecule is modeled as a point-like two-level system with the electric dipole set to perpendicular with the surface of the NP[9,25]. The matrix elements of electric dipole, magnetic dipole, and electric quadrupole momentum operators of the molecule are set as $\boldsymbol{\mu}_{12} = |e|(2\text{\AA},0,0)$, $\boldsymbol{m}_{21} = i|e|\omega_0(0.05\text{\AA}^2,0,0)$, $\Theta_{21,xx} = |e|(0.0625\text{\AA}^2)$, $\Theta_{21,yy} = -|e|(0.025\text{\AA}^2)$, $\Theta_{21,zz} = -|e|(0.0375\text{\AA}^2)$, $\Theta_{21,xy} = \Theta_{21,yx} = |e|(0.0625\text{\AA}^2)$, $\Theta_{21,xz} = \Theta_{21,zx} = |e|(0.03125\text{\AA}^2)$, $\Theta_{21,yz} = \Theta_{21,zy} = |e|(0.04375\text{\AA}^2)$.

The resonance wavelength $\lambda_0$ $(2\pi c/\omega_0)$ and the matrix element of relaxation operator $\gamma_{21}$ of the molecule are taken as 300 nm and 0.3eV, respectively. The CD signal of the system can be divided into two parts: $\text{CD} = \text{CD}(\text{ED-MD}) + \text{CD}(\text{ED-EQ})$, where $\text{CD}(\text{ED-MD})$ is caused by the interaction between the ED and the MD of the chiral molecule, and has been extensively discussed in the previous works[8,9,14]. The $\text{CD}(\text{ED-EQ})$ represents the contribution of the EQ, which has been usually ignored. Here we present an exact T-matrix method to calculate the CD of the system when the electric quadrupole contribution is considered. Details of the calculated process are given in the Methods section. The correctness of this method has been verified using the coupled multipole theory (see Supplementary S1)[26]. Fig. 1a shows the calculated results for the CD as a function of the wavelength. For the refraction index of Ag, Palik's data has been used[27], and the dielectric function of the embedded medium of the system is taken as $\varepsilon_m = 1.8$ (water). The results have been normalized by a factor of $\text{NA}/(0.23 I_{inc})$, where NA is the Avogadro constant and $I_{inc}$ is the intensity of the incident wave.

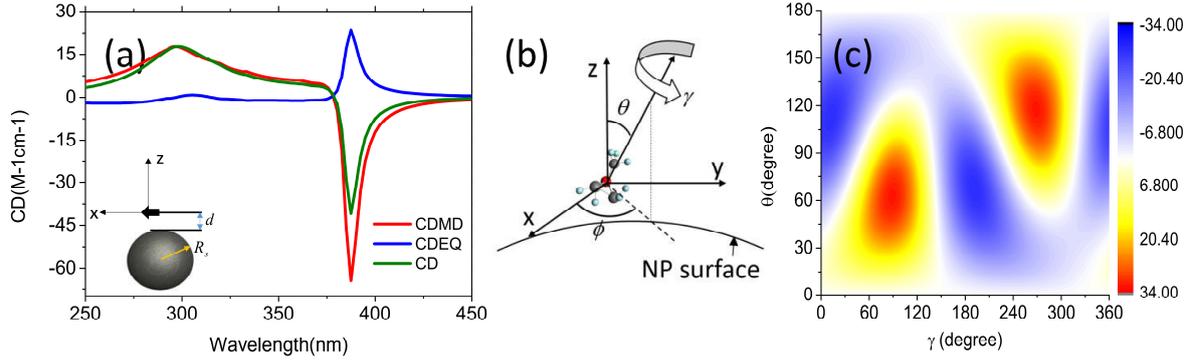

**Figure 1 | (a) Calculated CD signals for molecule-NP complex with chiral molecule put d=1 nm away from the surface of the Ag NP with $R_s = 10 \text{ nm}$. The parameters of the molecule are set as: $\boldsymbol{\mu}_{12} = e(2\text{Å},0,0)$, $\boldsymbol{m}_{21} = ie\omega_0(0.05\text{Å}^2,0,0)$, $\Theta_{21,xx} = e(0.0625\text{Å}^2)$, $\Theta_{21,yy} = -e(0.025\text{Å}^2)$, $\Theta_{21,zz} = -e(0.0375\text{Å}^2)$, $\Theta_{21,xy} = \Theta_{21,yx} = e(0.0625\text{Å}^2)$, $\Theta_{21,xz} = \Theta_{21,zx} = e(0.03125\text{Å}^2)$, $\Theta_{21,yz} = \Theta_{21,zy} = e(0.04375\text{Å}^2)$. Inset: Schematics of the system and the coordinate system. (b) Direction angle $\theta$, azimuth angle $\phi$, and rotational angle $\gamma$ specifying the orientation of the molecule with respect to the surface of the NP. (c) CD(ED-EQ) as a function of $\gamma$ and $\theta$.**

The red line and blue line correspond to $\text{CD}(\text{ED-MD})$ and $\text{CD}(\text{ED-EQ})$, respectively. It can be seen clearly that both $\text{CD}(\text{ED-MD})$ and $\text{CD}(\text{ED-EQ})$ exhibit peaks near 300 nm and 380 nm, which correspond to the molecule and plasmon resonance frequencies, respectively.

Although the $\text{CD}(\text{ED-EQ})$ is very small at the molecule resonance frequency, its contribution to the total CD (green line) is roughly equivalent comparing with that of the $\text{CD}(\text{ED-MD})$ at the plasmon resonance frequency. That is to say, the contribution of the EQ can not be ignored in such a case.

The $\text{CD}(\text{ED-EQ})$ depends critically on the direction of the molecule with respect to the surface of the NP. In Fig. 1c, we plot $\text{CD}(\text{ED-EQ})$ as functions of the molecular orientation angle $\theta$ and $\gamma$ with respect to the surface of the NP at $\lambda = 310 \text{ nm}$. The molecular orientation angles are specified by $\theta$, $\phi$ and $\gamma$ as shown in Fig. 1b. Here, $\theta$, $\phi$ are the direction angle and the azimuth angle of the molecular electric dipole momentum $\hat{\boldsymbol{\mu}}$, respectively. $\gamma$ denotes the rotation angle of the molecule about its electric dipole momentum. The multipole momenta of the molecule at $(\theta,\phi,\gamma)=(0,0,0)$ are taken to be

$\boldsymbol{\mu}_{12}=|e|(0,0,2\text{\AA})$, $\Theta_{21,xx}=-|e|(0.025\text{\AA}^2)$, $\Theta_{21,yy}=-|e|(0.0375\text{\AA}^2)$, $\Theta_{21,zz}=|e|(0.0625\text{\AA}^2)$,

$\Theta_{21,xy}=\Theta_{21,yx}=|e|(0.04375\text{\AA}^2)$, $\Theta_{21,xz}=\Theta_{21,zx}=|e|(0.0625\text{\AA}^2)$, $\Theta_{21,yz}=\Theta_{21,zy}=|e|(0.03125\text{\AA}^2)$.

In addition to some special angles such as $\theta=0°$ or $\theta=180°$, the contribution of the EQ to the total CD signal can always reach a considerable value. Practically, $\text{CD}(\text{ED-EQ})$ is most intense and reaches a value of 34 when $\theta=0°$ and $\gamma=90°$. Furthermore, with the varying of $\theta$, $\text{CD}(\text{ED-EQ})$ changes sign, which means that an altering in the direction of the molecular electric dipole momentum may lead to an inversion in the sign of CD signal. This may explain a previous experimental result which shows the CD at plasmon resonance wavelength may be inverted by altering the orientation of chiral molecules[13]. Here it's worth to mention that, if an average is performed over $\theta$ and $\gamma$, the signal is vanished. Thus, the EQ gives a contribution to the CD signal only when the nonrandom molecular alignment with respect to the surface of the NP appears[3,28-29]. On the other hand, in some cases such as the chiral molecule in hot spots of the plasmonic structure, it can dominate the optical activity phenomenon, and giant enhanced CD signal can be observed.

**Giant chiroptical effect by a Ag dimer.** We consider a chiral molecule being put in the hotspot of an Ag dimer as shown in Fig. 2a. Here the radii of two silver spheres are taken as 15 nm and the separation between them is $\text{d}=1\,\text{nm}$. The rotation axis of the Ag dimer is perpendicular with the electric dipole moment of the molecule. The parameters of the molecule are taken to be identical with those in Fig. 1a. The blue line and red line in Fig. 2b correspond to CD(ED-EQ) and CD(ED-MD), respectively. Comparing them with those in Fig. 1a, one can find that the EQ induced CD (CD(ED-EQ)) is greatly enhanced, with two positive peaks at 378 nm and 398 nm and two negative peaks at 387 nm and 434 nm. At $\lambda = 398$ nm, the value of CD(ED-EQ) can be as large as 9000 which is about 1000 times of the CD(ED-MD) at the peak. In addition, the angle dependence of CD(ED-EQ) is similar to the case of the single sphere as shown in Fig. 1c. In the large scale molecular orientation, the contribution of the electric quadrupole to the total CD signal plays a key role.

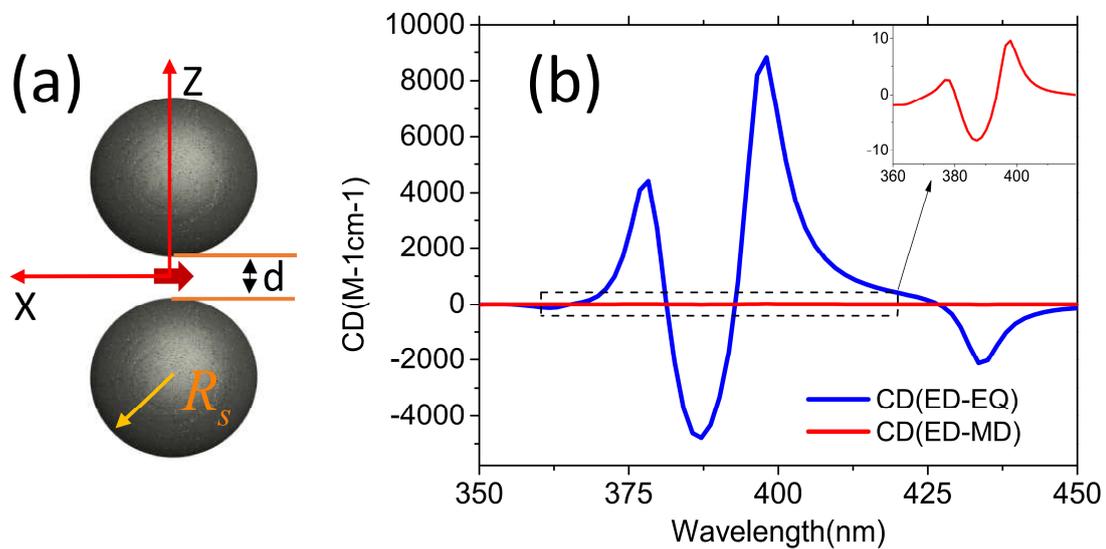

**Figure 2 | (a) Schematics of the considered system consisting of a Ag dimer and a chiral molecule. The radii of the NPs are set as $R_s = 15$ nm. The parameters of the molecule are taken to be same with those in Figure 1(a). (b) CD(ED-EQ) and CD(ED-MD) for the system as a function of wavelength. Inset: CD(ED-MD) from 360-400 nm.**

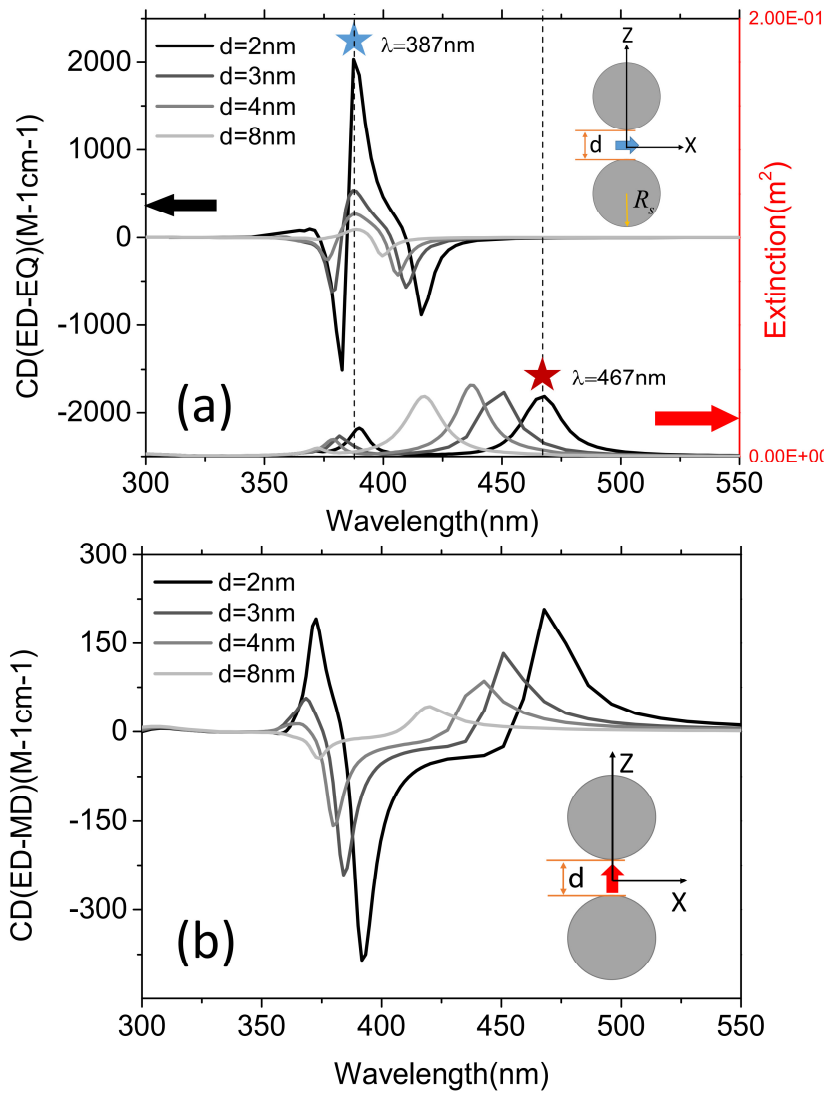

**Figure 3 | (a) CD(ED-EQ) signals for molecule inserted into Ag dimer with $R_s = 15\,\text{nm}$ at various inter particle distances d. Parameters of the molecule are set to be equal with those in Figure 2. (b) The corresponding results for CD(ED-MD) with the electric dipole of the molecule aligning with the connection line of the dimer.**

The results in Fig.2 are only for the case with $d = 1\,\text{nm}$. In fact, the EQ induced CD strongly depends on the separation distance between two NPs. Fig. 3a shows calculated $CD(ED\text{-}EQ)$ as a function of wavelength for various separation distances between two particles. The radii of Ag NPs are taken as 15nm, while parameters of the inserted molecule are identical with those in Fig. 2. The solid line, dashed line, dotted line and dot-dashed line correspond to the cases with $d = 2\,nm$, $3\,nm$, $4\,nm$ and $8\,nm$, respectively. For comparison, the corresponding extinctions are also given in the figure. It is seen clearly that the $CD(ED\text{-}EQ)$ increases rapidly with the decrease of inter particle distance. For example, when $d = 2\,nm$, the peak value of $CD(ED\text{-}EQ)$ is about 4 times larger than that with $d = 3\,nm$.

It is known that two peaks can be observed from the extinction spectra of the dimer, which come from two kinds of plasmon resonance. The peak at 380 nm (red solid line in Fig.3(a)) originates from the single scattering localized surface plasmon resonance of Ag NP, and does not shift a lot with the decrease of the gap. On the other hand, the peak in the longer wavelength region (467 nm when d=2 nm) undergoes a red-shift when the inter particle distance decreases, which is caused by coupling plasmon resonance. Comparing the extinction spectra with the $CD(ED\text{-}EQ)$, we find that strong $CD(ED\text{-}EQ)$ can only be observed in short wavelength region. It keeps in very small value in any case to change the gap of the Ag dimer. This is in contrast to the $CD(ED\text{-}MD)$ which has been investigated in previous works. For comparison, in Fig. 3b we plot the corresponding $CD(ED\text{-}MD)$ signals. Here the electric dipole momentum of the molecule is take as $\boldsymbol{\mu}_{12} = |e|(0,0,2\text{Å})$ and $\boldsymbol{m}_{21} = i|e|\omega_0(0,0,0.05\text{Å}^2)$. Due to the axial symmetry of the Ag dimer, the molecular quadrupole does not contribute to the CD in such a case. As can be seen, the peak of the $CD(ED\text{-}MD)$ at the long wavelength region appears, which corresponds to the coupling resonance peak in the extinction spectrum.

In order to disclose the physical origin of the phenomena, we calculate the electric fields and electric field gradients at the molecule position for the dimers discussed in Fig. 3. Fig. 4a and b show the electric field enhancement factor $|\boldsymbol{E}_{inc} + \boldsymbol{E}_s|/|\boldsymbol{E}_{inc}|$ as a function of wavelength, while the electric field gradient improvements $\sqrt{\sum_{ij}|\partial_i E_{incj}+\partial_i E_{sj}|^2}/\sqrt{\sum_{ij}|\partial E_{incj}|^2}$ are described by Fig. 4c and d. The electric polarizations of the incident wave are set to be parallel with the symmetry axis of the dimer system for Fig. 4a and c, while Fig 4b and d correspond to the perpendicular cases. From Fig. 4a, it is seen that the electric fields are strongly enhanced by the dimer. When the inter particle distance is set to be 2 nm, the enhancement factor can be as large as 300 at the plasmon resonance frequency (467 nm). Due to the existence of such a large electric field, $CD(ED\text{-}MD)$ is significantly enhanced, as shown in Fig. 3b. In contrast, the $CD(ED\text{-}EQ)$ is not improved much in such a case. This is because the electric field gradient is not large in such a case (see red star mark in Fig.4c). According to the theory (see Method section), $CD(ED\text{-}EQ)$ not only depends on the electric filed but also the electric field gradient. At the single scattering resonance frequency such as $\lambda \sim 387$ nm for d= 2 nm (marked by blue stars in Fig. 4), both the electric field and its gradient are large, the $CD(ED\text{-}EQ)$ is improved strongly.

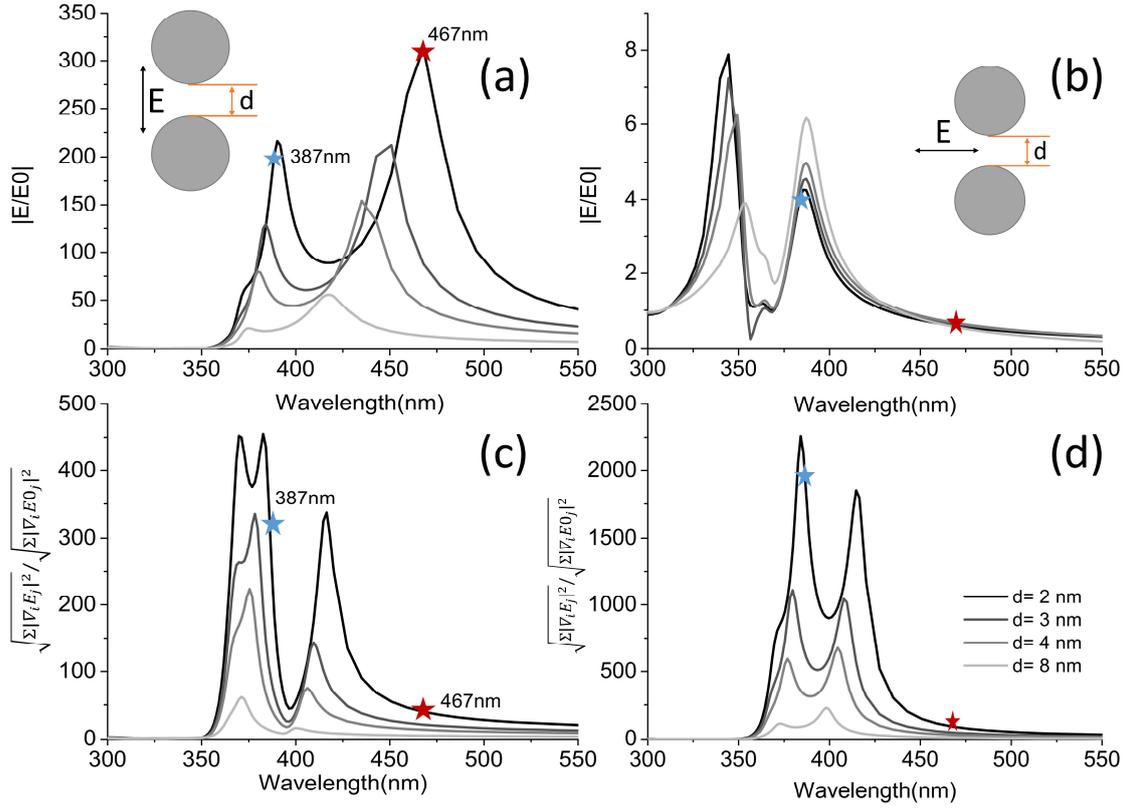

**Figure 4 | (a,b)** Enhancement factors $|E_{inc} + E_s|/|E_{inc}|$ as a function of wavelength for the electric field in the center of the Ag dimer discussed in Figure 3. **(c,d)** Electric field gradient enhancement factors $\sqrt{\sum_{ij}|\partial_i E_{inc\,j} + \partial_i E_{s\,j}|^2} / \sqrt{\sum_{i}|\partial E_{inc\,j}|^2}$ versus the wavelength of incident light. The electric field of the incident wave is parallel with the connection line of the dimer for (a, c), and are perpendicular for (b,d).

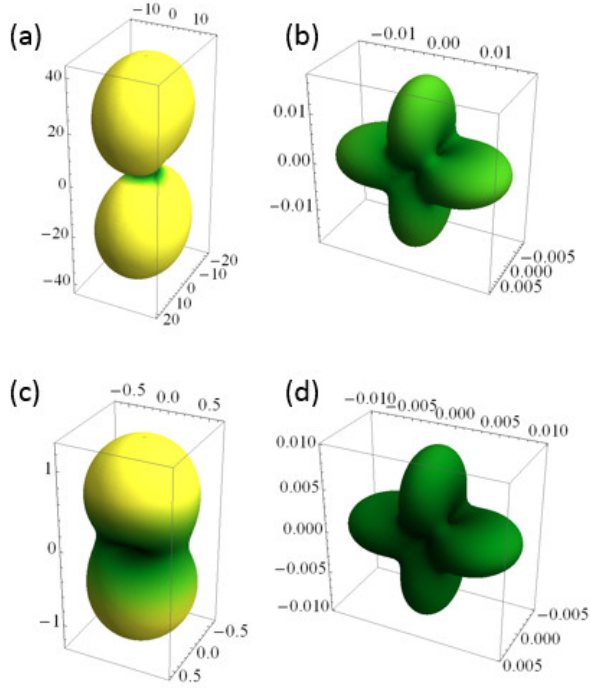

**Figure 5 | Far field radiation patterns** $\lim_{r\to\infty}\left|rE^{(E2)}\right|$ **for an electric quadrupole with** $\Theta_{xz}=\Theta_{zx}=1$. **(a) and (c) represent the results for the Ag dimer discussed in Figure 4 with inter particle distance** $d=2\,\text{nm}$ **when the quadrupole is inserted into the gap. (b) and (d) represent the results without the Ag dimer. The electric quadrupole radiates at the wavelength of 387 nm for (a) and (b); 467 nm for (c) and (d).**

Previous investigations have shown that dissipative chiral currents inside the mental, giving arise to a large $CD(ED\text{-}MD)$ from the NP, can be induced by the electric dipole (E1) of the molecule.[8-9,14] Here according to the theory (see Method section for detail), the dissipative chiral currents can be also generated by the *induced* electric quadrupole (E2). The amount of chiral currents inside the NP is proportional to the emission intensity of the electric quadrupole [18, 20, 21]. Thus, the $CD(ED\text{-}EQ)$ is related to the electric quadrupole emission of the molecule. In Fig. 5a, we plot the angular distributions of the scattering field amplitude $\lim_{r\to\infty}\left|rE^{(E2)}\right|$ for an electric quadrupole inserted into the hotspot of the Ag dimer discussed in Fig. (4). The gap of the dimer system is set as $d=2\,\text{nm}$, and elements of the electric quadrupole is given as $Q_{xz}=Q_{zx}=1$ with other elements set to be null. The wavelength of the quadrupole radiation is 387 nm, which corresponds to the single scattering resonance wavelength of the dimer. The corresponding results for the electric quadrupole radiation in the absence of the nano dimer is presented in Fig. 5b. Comparing them, one can observe the emission of the electric quadrupole is improved by a factor of ~ 2000. On the other hand, if the wavelength of the quadrupole radiation is set as to be equal with the coupling resonance wavelength of the dimer, the emission enhancement factor diminishes dramatically. Fig. 5c and d are arranged the same as Fig. 5a and b, except the wavelength of radiation is set as 467 nm. Comparing Fig. 5c with Fig. 5d, we find the enhancement factor only reaches ~50, which is much smaller than that at 387 nm. This can be seen as another point of view to reveal the origin of the phenomena as has been shown in Fig. 3.

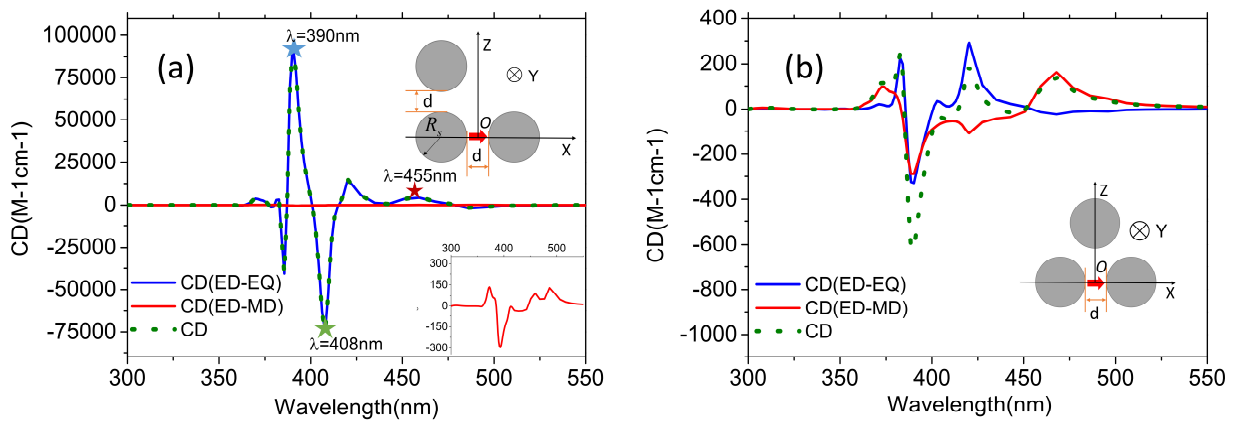

**Figure 6 |** (a) CD spectrum for molecule inserted into a right-angle trimer system. The radii of Ag NPs are 15 nm with inter particle distance d= 2 nm. The molecule is inserted into a gap of two NPs with electric dipole aligning with the connection line. (b) Corresponding results for molecule inserted into an equilateral trimer system. Distances d between NPs are 2 nm. The parameters of the Ag NPs and the molecule are same with (a).

**Giant chiroptical effect in trimers.** In addition to the above mentioned factors, $CD(ED\text{-}EQ)$ is also extremely sensitive to the symmetry of the nanostructure. When the electric dipole of the molecule at the hotspots of the dimer is taken to be parallel with the symmetry axis as shown in Fig. 3b, the CD signal is only determined by $CD(ED\text{-}MD)$ and the contribution of $CD(ED\text{-}EQ)$ is zero. However, when another NP is introduced into the dimer system[24], the situation becomes different. If the third NP is put on the top side of a particle to form a right-angle trimer as shown in the inset of Fig. 6a, extremely giant signal caused by the electric quadrupole of the molecule can be observed. The blue solid line in Fig. 6a represents such a case. Here the radii of three silver particles are taken to be 15 nm and the distances between them are 2 nm. The parameters of the molecule are taken to be equal with those in Fig. 1a.

From Fig. 6a, we find that the $CD(ED\text{-}EQ)$ at 390 nm reaches 100000 which is ~300 times greater than that of $CD(ED\text{-}MD)$ (the red line). The total CD signal (the green dotted line) is utterly dominated by the $CD(ED\text{-}EQ)$. Comparing Fig. 6a with Fig. 3a, we find the maximum of the $CD(ED\text{-}EQ)$ for this case is 50 times larger than that in Fig. 3a, when the parameters of the molecule are taken as the same. This huge $CD(ED\text{-}EQ)$ signal can be attributed to the following two aspects: (1) The electric field and the gradient of the electric field at the molecule position are enlarged by the symmetry breaking caused by the third NP; (2) The radiations for molecule E1 and E2 are enhanced simultaneously by introducing the third NP. Such a phenomenon depends on the symmetry of the structure. For example, if we change the position of the third particle and construct an equilateral trimer as shown in the inset of Fig. 6b, the huge enhanced effect of the $CD(ED\text{-}EQ)$ does not appear. In such a case, the calculated results for $CD(ED\text{-}EQ)$ and $CD(ED\text{-}MD)$ are given in Fig. 6b.

The blue line and red line in Fig. 6b correspond to the $CD(ED\text{-}EQ)$ and $CD(ED\text{-}MD)$, respectively, the green dotted line to the total CD. It is shown clearly that the peak value of $CD(ED\text{-}EQ)$ is comparable with that of the $CD(ED\text{-}MD)$. In the long wavelength region the $CD(ED\text{-}EQ)$ is completely suppressed by the $CD(ED\text{-}MD)$. To disclose the mechanism behind the phenomenon, we decompose $\mathrm{CD}(\mathrm{ED\text{-}EQ})$ into signals caused by the different components of the electric quadrupole momentum of the molecules. It is found that the extremely large signal in the right-angle trimer system is dominated by the *ZY* and *YZ* components of the molecular electric quadrupole. According to the symmetry analysis (see Method section for detail), we can prove the absorption rates caused by these two terms are the same for left- and right handed waves, when the mental cluster possesses a mirror symmetry about the YZ plane as shown in the inset of Fig. 6b. Thus, the $\mathrm{CD}(\mathrm{ED\text{-}EQ})$ for the system described in Fig. 6b is not enhanced as much as that in Fig. 6a.

**Discussion**

The above calculations only focus on the hybrid system consisting of a chiral molecule and spherical NPs. In fact, the method provided in this work is applicable to the CD calculation of chiral molecules in a finite cluster with arbitrarily disposed objects, the phenomena disclosed in this work can also be found in some nanocomposites consisting of a chiral molecule and non-spherical NPs. In Supplementary Information (see Supplementary S2) we present the calculated results for the $\mathrm{CD}(\mathrm{ED\text{-}MD})$ and $\mathrm{CD}(\mathrm{ED\text{-}EQ})$ in a nanocomposite consisting of a chiral molecule and a cylindrical silver nanoparticle. In such a case, the CD signals also depend on the relative position between the chiral molecule and the nanoparticle. We found that when the molecule is put in the vicinity of the edge of the silver cylinder, the $\mathrm{CD}(\mathrm{ED\text{-}EQ})$ can be much larger than the $\mathrm{CD}(\mathrm{ED\text{-}MD})$ in many regions of the CD spectrum, which is in contrast to the case of single sphere system shown in Fig.1. This is due to the cylindrical object possesses sharp edges where electric charges can accumulate and produce much larger electric field gradient enhancements.

Based on the above understanding, we can design the nanostructures to observe giant chiroptical effects using the EQ contribution at the required frequency, which are very beneficial for ultrasensitive detection and characterization of biomolecules. In addition, an unexpectedly large CD response in the nanocomposite consisting of a chiral molecule and colloidal silver nanocubes, which has been observed in recent experiment, can be well understood if the contribution of the EQ is considered. Due to the existence of sharp edges and apexes, multipole plasmon modes can be excited, giving arise to highly nonuniform electric fields on the surface of the nano cubes. Thus, the enhancement factor for the electric field gradient is supposed to be much larger than that of the electric field, which results in that the interaction of the EQ with light is stronger than that of the ED. Furthermore, according to the previous investigations [4,12,13], the strong CD signals occur only in plasomon modes involving higher order plasmon modes rather in the bright dipole mode. This also suggests that the large signal are very likely to be induced by the EQ emission of the chiral molecule (E2), which can excite the dark (higher order) plasmon mode more efficiently than the electric dipole emission of the molecule (E1). That is to say, the contribution of $CD(ED\text{-}EQ)$ may dominate the total CD signal in these experiments. At last, we would like to point out that for separation d >1 nm the influence of quantum effects is very small, so that our plasmonic subsystem can be described by the classical theory of the local dielectric function and is within the classical regime[30].

In summary, we have presented an exact T-matrix method to calculate the CD of molecule−NP nanocomposites with the electric quadrupole contribution. We have studied the EQ contribution to the plasmon-induced CD signals for various nanocomposites such as chiral molecules being put in the vicinity of single NP or inserted into hotspots of nano dimer or trimer system. We have found that the electric quadrupolar contribution can play a key role in many cases. For example, when the rotational symmetry of the nano dimer is broken by the presence of a third nano particle, CD signals caused by the electric quadrupole can be 300 times larger than those predicted by the dipole theory. The physical origin for such a phenomenon has been disclosed. This means that we can design nanostructures to observe giant chiroptical effects according to the requirements, which would be helpful for the experimental design of novel biosensors to realize ultrasensitive probing of chiral information on molecules by plasmon-based nanotechnology.

**Methods**

**The T-matrix method for calculating CD(ED-MD) and CD(ED-EQ).** In this section, we provide the calculated method. Consider a semi-classical hybrid system consisting of a molecule and a cluster of N NPs, which is excited by circularly polarized light. The molecule (situated at the origin of the coordinate $\mathbf{r}_{mol}=\mathbf{0}$) is assumed to be point-like two-level systems, no vibrational structure of transitions is considered. When electric quadrupole terms are neglected, the calculated method has been given in one of our previous works[25]. In the following, we present a method to calculate the CD signal with the contribution of electric quadrupole terms. The master equation for quantum states of the molecule can be written as

$$\hbar \frac{\partial \rho_{ij}}{\partial t} = i\left[\rho, \hat{H}_0 + \hat{H}_I\right]_{ij} - \Gamma_{ij}(\rho) \quad (1)$$

where $\hat{H}_0$ is the unperturbed Hamiltonian of the molecule, $\rho$ is the density matrix. $\hat{\Gamma}_{ij}$ denotes the relaxation term which describes the phenomenological damping, and for the two level quantum system, it can be expressed as $\Gamma_{11}(\rho) = -\Gamma_{22}(\rho) = -\gamma_{22}\rho_{22}$, $\Gamma_{12}(\rho) = \gamma_{21}\rho_{12}$, $\Gamma_{21}(\rho) = \gamma_{21}\rho_{21}$, with $\rho_{ij}$ representing the element of the density matrix. $\hat{H}_I = -\hat{\boldsymbol{\mu}}\cdot\boldsymbol{E}_T - \hat{\boldsymbol{m}}\cdot\boldsymbol{B}_T - \frac{1}{3}\hat{\Theta}:\nabla E$ is the light-mater interaction operator ( : is the operator which denotes tensor contraction), with $\hat{\boldsymbol{\mu}}$, $\hat{\boldsymbol{m}}$ and $\hat{\Theta}$ denoting the electric dipole, magnetic dipole and electric quadrupole operators, respectively. $(\boldsymbol{E}_T, \boldsymbol{B}_T) = \text{Re}\left[(\boldsymbol{E}_{tot}, \boldsymbol{B}_{tot})e^{-i\omega t}\right]$ are the EM field incident on the molecule. The master equation can be solved using rotating-wave approximation, and absorption of the system is given by

$$Q = Q_{mol} + Q_{NP}. \tag{2}$$

The $Q_{mol}$ denotes the absorption rate of the molecule, which is given by

$$Q_{mol} = \frac{\omega_0 \gamma_{21}}{2} \frac{\left| \boldsymbol{\mu}_{21} \cdot \boldsymbol{E}_{tot}^{(0)}(\boldsymbol{r}_{mol}) + \boldsymbol{m}_{21} \cdot \boldsymbol{B}_{tot}^{(0)}(\boldsymbol{r}_{mol}) + \frac{1}{3} \Theta_{21} : \nabla \boldsymbol{E}_{tot}^{(0)}(\boldsymbol{r}_{mol}) \right|^2}{\left| \hbar(\omega - \omega_0) + i\gamma_{21} - G \right|^2} \tag{3}$$

where $\boldsymbol{\mu}_{21} = \langle 2 | \hat{\boldsymbol{\mu}} | 1 \rangle$, $\boldsymbol{m}_{21} = \langle 2 | \hat{\boldsymbol{m}} | 1 \rangle$ and $\Theta_{21} = \langle 2 | \hat{\Theta} | 1 \rangle$ represent the matrix elements of $\hat{\boldsymbol{\mu}}$, $\hat{\boldsymbol{m}}$ and $\hat{\Theta}$, respectively. $G$ is the function which describes the broadening of the resonance peak of the signal caused by the interactions between the molecules and NPs, and its expression can be founded elsewhere[14]. $\omega_0$ is the frequency of molecular transition; $\boldsymbol{r}_{mol}$ is the position vector of the molecule. $\left( \boldsymbol{E}_{tot}^{(0)}, \boldsymbol{B}_{tot}^{(0)} \right) = \left( \boldsymbol{E}_{inc}^{(0)} + \boldsymbol{E}_s^{(0)}, \boldsymbol{B}_{inc}^{(0)} + \boldsymbol{B}_s^{(0)} \right)$, with $\left( \boldsymbol{E}_{inc}^{(0)}, \boldsymbol{B}_{inc}^{(0)} \right)$ being the electric and magnetic fields of the incident plane wave. $\left( \boldsymbol{E}_s^{(0)}, \boldsymbol{B}_s^{(0)} \right)$ is the EM field scattered by the NPs in the absence of the molecule.

$Q_{NP}$ is the absorption contributed by the NPs, which can be expressed as

$$Q_{NP} = -\frac{1}{2} \mathrm{Re} \sum_{\xi=1}^{N} \oiint_{\Omega_\xi} \boldsymbol{n} \cdot \boldsymbol{E}_{tot}(\boldsymbol{r} - \boldsymbol{r}_\xi) \times \boldsymbol{H}_{tot}^*(\boldsymbol{r} - \boldsymbol{r}_\xi) d\Omega_\xi, \tag{4}$$

where $\boldsymbol{r}_\xi$ denotes the position vector of the $\xi^{th}$ sphere. $\Omega_\xi$ represents a surface circumscribing around the $\xi^{th}$ NP. The complex electric $\boldsymbol{E}_{tot}(\boldsymbol{r} - \boldsymbol{r}_\xi)$ and magnetic fields $\boldsymbol{H}_{tot}(\boldsymbol{r} - \boldsymbol{r}_\xi)$ in space can be written as

$$\begin{aligned} \boldsymbol{E}_{tot} &= \boldsymbol{E}_{tot}^{(E1)} + \boldsymbol{E}_{tot}^{(M1)} + \boldsymbol{E}_{tot}^{(E2)} + \boldsymbol{E}_{tot}^{(0)} \\ \boldsymbol{H}_{tot} &= \boldsymbol{H}_{tot}^{(E1)} + \boldsymbol{H}_{tot}^{(M1)} + \boldsymbol{H}_{tot}^{(E2)} + \boldsymbol{H}_{tot}^{(0)} \end{aligned} \tag{5}$$

where $\left(\boldsymbol{E}_{tot}^{(E1)}, \boldsymbol{H}_{tot}^{(E1)}\right)$, $\left(\boldsymbol{E}_{tot}^{(M1)}, \boldsymbol{H}_{tot}^{(M1)}\right)$, $\left(\boldsymbol{E}_{tot}^{(E2)}, \boldsymbol{H}_{tot}^{(M2)}\right)$ stand for the electric and magnetic fields caused by $\boldsymbol{\mu}^{(E1)}$, $\boldsymbol{m}^{(M1)}$, $\Theta^{(E2)}$. The $\boldsymbol{\mu}^{(E1)} = \mathrm{Tr}(\rho\hat{\boldsymbol{\mu}})$, $\boldsymbol{m}^{(E1)} = \mathrm{Tr}(\rho\hat{\boldsymbol{m}})$ and $\Theta^{(E2)} = \mathrm{Tr}(\rho\hat{\Theta})$ are called the *induced* electric dipole, the *induced* magnetic dipole, and the *induced* electric quadrupole of molecule. The components of the $\boldsymbol{\mu}^{(E1)}$, $\boldsymbol{m}^{(E1)}$, $\Theta^{(E2)}$ are given by

$$\mu_a^{(E1)} \approx \tilde{\alpha}_{ab} E_{tot\,b}^{(0)} + \tilde{G}_{ab} B_{tot\,b}^{(0)} + \frac{1}{3}\tilde{A}_{a,bc}\partial_b E_{tot\,c}^{(0)} \tag{6}$$

$$m_a^{(M1)} \approx -\tilde{G}_{ba} E_{tot\,b}^{(0)} \tag{7}$$

$$\Theta_{ab}^{(E2)} \approx \tilde{A}_{c,ab} E_{tot\,c}^{(0)} \tag{8}$$

in the Cartesian coordinate. The $\tilde{\alpha}_{ab}$, $\tilde{G}_{ab(ba)}$, and $\tilde{A}_{a(c),bc(cb)}$ represent elements of dynamic molecular response tensors, which can be expressed as

$$\begin{aligned}\tilde{\alpha}_{ab} &= -\frac{1}{\hbar(\omega-\omega_0)+i\gamma_{21}-G}(\boldsymbol{\mu}_{12}\otimes\boldsymbol{\mu}_{21})_{ab} \\ \tilde{G}_{ab} &= -\frac{1}{\hbar(\omega-\omega_0)+i\gamma_{21}-G}(\boldsymbol{\mu}_{12}\otimes\boldsymbol{m}_{21})_{ab} \\ \tilde{A}_{a,bc} &= -\frac{1}{\hbar(\omega-\omega_0)+i\gamma_{21}-G}(\boldsymbol{\mu}_{21})_a(\Theta_{21})_{bc}\end{aligned} \tag{9}$$

Inserting equation (5) into equation (4), the absorption contributed by the NPs can be rewritten as

$$Q_{NP} = Q_{NP}^{(0-0)} + Q_{NP}^{(0-M1)}(\tilde{G}) + Q_{NP}^{(0-E2)}(\tilde{A}) + Q_{NP}^{(0-E1)}(\tilde{\alpha},\tilde{G},\tilde{A}) + Q_{NP}^{(E1-E1)}(\tilde{\alpha},\tilde{G},\tilde{A}) \tag{10}$$

where

$$Q_{NP}^{(X1-X2)} = -\frac{1}{2(1+\delta(X1,X2))}\mathrm{Re}\sum_{\xi=1}^{N}\iint_{\Omega_\xi}\boldsymbol{n}\cdot\left[\boldsymbol{E}_{tot}^{(X1)}(\boldsymbol{r}-\boldsymbol{r}_\xi)\times\boldsymbol{H}_{tot}^{(X2)*}(\boldsymbol{r}-\boldsymbol{r}_\xi)+\boldsymbol{E}_{tot}^{(X2)}(\boldsymbol{r}-\boldsymbol{r}_\xi)\times\boldsymbol{H}_{tot}^{(X1)*}(\boldsymbol{r}-\boldsymbol{r}_\xi)\right]d\Omega_\xi \tag{11}$$

$\delta(X1,X2) = 1$ when $X1$ is same with $X2$, and $\delta(X1,X2) = 0$ if $X1$ and $X2$ are different.

The CD signal is defined as the difference between absorption of left- and right- handed polarized wave, which can be expressed as

$$\mathrm{CD} = \langle Q^+ - Q^- \rangle_\Omega \qquad (12)$$

where $\langle \ \rangle_\Omega$ denotes average is taken over the solid angle of the incident field. This averaging is necessary because the systems are usually randomly oriented in solution. In fact, from equations 3 and 10, CD mainly comes from four aspects of contributions: field-filed interaction CD(0-0), electric dipole-electric dipole interaction CD(ED-ED), electric dipole-magnetic dipole interaction CD(ED-MD), and electric dipole- electric quadrupole interaction CD(ED-EQ), which are determined by the products of $\mu_{12(21)}$, $m_{12(21)}$ and $Q_{12(21)}$. Here, the terminologies 'electric dipole', 'magnetic dipole', 'electric quadrupole' represent the matrix elements of multipole operators and should not be confused with the *induced* multipole moments $\mu^{(E1)}$, $m^{(E1)}$, $\Theta^{(E2)}$ discussed above. If the cluster formed by the nano particles is achiral, CD(0-0) will be vanished. Furthermore, if $\mu$ is in the plane formed by the centers of the NPs which is exactly same with the cases discussed in this article, CD(ED-ED) also equals to zero. The CD signal is only contributed by CD(ED-MD) and CD(ED-EQ). Thus, the total signal can be written as

$$\mathrm{CD} = \mathrm{CD}(\mathrm{ED\text{-}MD}) + \mathrm{CD}(\mathrm{ED\text{-}EQ}). \qquad (13)$$

According to the T-matrix (see Supplementary S3), CD can be written as

$$\mathrm{CD}(X) = 2\sum_{v=1}^{\infty} \mathrm{Re}\left\{ \left[ \langle M(X) \rangle_\Omega \right]_{wv}^{12} a_v^{\sigma+} b_v^{\sigma+*} + \left[ \langle M(X) \rangle_\Omega \right]_{wv}^{21} b_v^{\sigma+} a_v^{\sigma+*} \right\} \qquad (14)$$

where $X$ stands for either ED-MD or ED-EQ, $M(X)$ is the light-mater interaction matrix determined by the property of the illuminated system, with $a_v^{\sigma+}$ and $b_v^{\sigma+}$ being the vector spherical function expansion coefficients of the left handed polarized wave. v = (m, n) is introduced here for notation simplification, and the convention v=1, 2, …, when n=1, 3, …, and m=-n,…n is used.

**Symmetry analysis on components of the molecular electric quadrupole momentum.**
Considering the electric dipole of the molecule is set to be align the X direction and the electric quadrupole moment of the molecule only has YZ and ZY components. From equations 3 and 10 and our numerical calculation, the large CD(ED-EQ) signal in Fig. 6a is proportional with the products of $\mu_{21x}$ and $\Theta_{yz}$ or $\Theta_{zy}$, which are invariant under mirror reflection about the YZ plane. Thus, when such a molecule (with only $\mu_{21x}$, $\Theta_{yz}$, $\Theta_{zy}$ components) is inserted into the gap of a structure like that in Fig. 6b, the following equations are established:

$$\begin{aligned}\left[\langle M(ED-EQ)\rangle_\Omega\right]^{12}_{vv} &= \left[\langle \tilde{M}(ED-EQ)\rangle_\Omega\right]^{12}_{vv} \\ \left[\langle M(ED-EQ)\rangle_\Omega\right]^{21}_{vv} &= \left[\langle \tilde{M}(ED-EQ)\rangle_\Omega\right]^{21}_{vv}\end{aligned} \quad (15)$$

where $\tilde{M}(ED-EQ)$ is the light-mater interaction matrix of the mirror reflected system which is oppositely handed. This is because, unlike the system discussed in Fig. 6a, the Ag nano cluster in Fig. 6b is also mirror symmetric about the YZ plane.

When two systems are oppositely handed with each other, their CD signals are related by

$$\text{CD} = -\text{CD}' \quad (16)$$

where CD and CD' are the signals which are caused by two systems which are enantiotopic. Inserting equations (14) and (15) into equation (16), the following equation is established:

$$\begin{aligned}&2\sum_{v=1}^{\infty}\text{Re}\left\{\left[\langle M(X)\rangle_\Omega\right]^{12}_{vv}a_v^{\sigma+}b_v^{\sigma+*} + \left[\langle M(X)\rangle_\Omega\right]^{21}_{vv}b_{v+}^{\sigma+}a_{v+}^{\sigma+*}\right\} \\ &= -2\sum_{v=1}^{\infty}\text{Re}\left\{\left[\langle \tilde{M}(X)\rangle_\Omega\right]^{12}_{vv}a_v^{\sigma+}b_v^{\sigma+*} + \left[\langle \tilde{M}(X)\rangle_\Omega\right]^{21}_{vv}b_{v+}^{\sigma+}a_{v+}^{\sigma+*}\right\}\end{aligned}. \quad (17)$$

In such a case, according to equation (14), it can be easily find that $\text{CD} = 0$.

**Acknowledgment**

This work was supported by the National Key Basic Research Special Foundation of China under Grant No. 2013CB632704 and the National Natural Science Foundation of China through Grants No. 61421001 and 11574030.

**Author contributions**

Numerical results and theoretical method are presented by T. W. with the help of W. Z, the idea and physical analysis are given by X. Z. and R.W. All authors reviewed the manuscript.

**Additional information:**

Supplementary information accompanies this paper at

http://www.nature.com/naturecommunications

**Competing financial interests**: The authors declare no competing financial interests.

Reprints and permission information is available online at

http://npg.nature.com/reprintsandpermissions/